\documentclass[a4paper,12pt]{article}

\usepackage{amsmath,amsthm,amssymb,bm} 
\usepackage{tikz} 

\newcommand{\oo}{\mathcal{O}} 
\newcommand{\ti}[1]{\tilde{#1}}     
\newcommand{\f}[2]{{\frac{#1}{#2}}} 
\def\s{\sqrt}
\newcommand{\sfrac}[2]{{#1/#2}}     
\def\l{\left}  
\def\r{\right} 
\newcommand{\fgref}[1]{{\figurename\;\ref{#1}}} 
\newcommand{\secref}[1]{{Section\;\ref{#1}}}   
\newcommand{\appref}[1]{{Appendix\;\ref{#1}}}   
\newcommand{\SU}{\mathrm{SU}} 
\newcommand{\sech}{\operatorname{sech}} 
\newcommand{\sn}{\operatorname{sn}} 
\newcommand{\cn}{\operatorname{cn}} 
\newcommand{\dn}{\operatorname{dn}} 
\newcommand{\ext}{\operatorname{ext}} 

\begin{document}

\begin{titlepage}
\thispagestyle{empty}
\begin{flushright}
IPMU14-0364,
UT-14-49,
RIKEN-QHP-176,
RIKEN-MP-105
\end{flushright}

\bigskip

\begin{center}
{\Large 
Holographic Entanglement 
and Causal Shadow 

\medskip

in 
Time-Dependent
Janus Black Hole

}
\end{center}

\bigskip

\begin{center}

Y\={u}ki {\sc Nakaguchi}$^{\heartsuit\clubsuit,}$%
\footnote{E-mail :  yuuki.nakaguchi@ipmu.jp},
Noriaki {\sc Ogawa}$^{\diamondsuit,}$%
\footnote{E-mail : noriaki@riken.jp}
and
Tomonori {\sc Ugajin}$^{\spadesuit,}$%
\footnote{E-mail : ugajin@kitp.ucsb.edu}

\setcounter{footnote}{0}

\bigskip

${}^\heartsuit$
{\it
Institute for the Physics and Mathematics of the Universe,
University of Tokyo, \\
Kashiwa, Chiba 277-8583, Japan
}\\

\medskip

${}^\clubsuit$
{\it
Department of Physics, Faculty of Science,
University of Tokyo, \\
Bunkyo-ku, Tokyo 133-0022, Japan
}\\

\medskip

${}^\diamondsuit$
{\it
Quantum Hadron Physics Laboratory \& Mathematical Physics Laboratory, \\
RIKEN Nishina Center, 
Wako, Saitama 351-0198, Japan
}\\

\medskip

${}^\spadesuit$
{\it
Kavli Institute for Theoretical Physics, University of California, \\
Santa Barbara, 
CA 93106, USA
}\\
\end{center}

\bigskip

\begin{abstract}
We holographically compute an inter-boundary entanglement entropy 
in a time-dependent two-sided black hole which was constructed 
in \cite{Bak:2007jm}
by applying time-dependent Janus deformation to BTZ black hole.
The black hole contains ``causal shadow region'' 
which is causally disconnected from both the conformal boundaries. 
We find that the Janus deformation results in an earlier phase transition 
between the extremal surfaces and that 
the phase transition disappears when the causal shadow is sufficiently large.
\end{abstract}

\end{titlepage}

\tableofcontents


\section{Introduction}

The relation between entanglement and black hole interior 
has attracted much attention recently
\cite{Mathur:2009hf,Almheiri:2012rt,Papadodimas:2012aq,Maldacena:2013xja}.
For eternal AdS black holes,
it was discussed that the time evolution of holographic entanglement entropy
\cite{Ryu:2006bv,Ryu:2006ef,Hubeny:2007xt}
can capture some information about the black hole interior,
taking a particular time slicing
with which the black hole looks time dependent. 
For subsystems composed of two disjoint same intervals located in each of two CFT's, the original CFT and the thermofield doubled copy CFT \cite{Hartman:2013qma},
its holographic entanglement entropy grows linearly in time for a while,
in accordance with the growth of the wormhole inside the black hole.
At a certain critical time,
the entropy becomes saturated
at twice the value of the black hole thermal entropy. 
In the dual CFT language,
this time dependent behavior of the entanglement entropy
is interpreted in terms of global quench process \cite{Calabrese:2005in}.
For such Calabrese Cardy type of two dimensional quenches, a systematic construction of their holographic duals was discussed in \cite{Ugajin:2013xxa}.

\vspace{0.1cm}

More general two-sided black holes can have even richer interior structures.
For example, similar inter-boundary entanglement entropies in
charged or rotating black hole geometries, which have vertically extended Penrose diagrams, 
were investigated in \cite{Caputa:2013eka,Iizuka:2014wfa}.
In this paper,
we focus on another interesting class of two-sided black holes
with a so called ``causal shadow'' region, 
which is a bulk region causally inaccessible from both the boundaries. 
The implications of such a region for holographic entanglement entropy
have been discussed \cite{Headrick:2014cta,Fischetti:2014uxa}.
For example, we can construct 
an asymptotically AdS black holes with a causal shadow
by sending shock waves from the boundaries
of eternal AdS black holes
\cite{Shenker:2013pqa,Shenker:2013yza,Shenker:2014cwa},
and we can also discuss its dual CFT \cite{Roberts:2014ifa}.
It is an interesting question how the dual CFT encodes 
information on causal shadow regions.

\vspace{0.1cm}

To investigate this question further,
we concentrate on another type of black hole with a causal shadow
called the three-dimensional time-dependent Janus black hole,\footnote{
There is also a static type of Janus deformation of BTZ black hole \cite{Bak:2011ga}.}
which is a one parameter deformation of the BTZ black hole and a solution of the Einstein-scalar theory \cite{Bak:2007jm}.
This black hole geometry has a nontrivial dilaton configuration,
without which it reduces to just the eternal BTZ black hole.
From the viewpoint of the dual boundary theory,
this nontrivial dilaton configuration corresponds to
the difference in the coupling constant and so in Hamiltonian between 
the two CFT's,
the original CFT and the thermofield doubled copy CFT \cite{Bak:2007jm}.
Its corresponding CFT state was proposed \cite{Bak:2007qw}
as a natural extension of
the usual eternal AdS black hole/thermofield double state correspondence
\cite{Maldacena:2001kr,Balasubramanian:1998de},
and this proposal was checked 
by computing a one point function
both on the CFT side and the gravity side \cite{Bak:2007jm,Bak:2007qw}.

\vspace{0.1cm}

In this paper,
we study the time evolution of
an inter-boundary holographic entanglement entropy
in the Janus black hole geometry,
expecting to capture some information on its causal shadow.
As in the BTZ black hole geometry,
there are two extremal surfaces for the subsystem we take,
where the entanglement entropy is given as the area of the one
with the smaller area. 
One which we call ``connected surface'' passes through the black hole interior
and connects the two asymptotic boundaries,
while the other which we call ``disconnected surface'' does not pass through the black hole interior but can penetrate partially into the interior.

\vspace{0.1cm}

In the Janus black hole geometry with not so large deformation parameter,
there is a critical time $t_c$ at which
the surface giving the entanglement entropy
switches from the connected one to the disconnected one,
as in the BTZ black hole geometry.
We find that the critical time $t_c$ is shorter than that in the BTZ case.
This is roughly because 
the deformation enlarges the wormhole region and so increases the connected surface area. 
In the black hole geometry with a sufficiently large deformation parameter, we find that
the area of the disconnected one becomes always  smaller
and that the holographic entanglement entropy 
is already proportional to the size of the subsystem from the beginning,
unlike the BTZ case.

\vspace{0.1cm}

This paper is organized as follows. 
In \secref{sec:properties}, 
we review properties of 
the three-dimensional Janus black hole
with emphasis on the difference 
from the BTZ black hole. 
In \secref{sec:timeevolution}, 
we compute the area of extremal surfaces with appropriate boundary conditions in this black hole geometry. 
In \secref{sec:phase_trans} 
we discuss the time evolution of the holographic entanglement entropy. 
We conclude this paper in \secref{sec:conc}.

\section{Properties of three-dimensional Janus Black Hole} \label{sec:properties}
Here we summarize the properties of 
the three dimensional time-dependent Janus black hole 
with emphasis on its causal structure and 
 dual CFT interpretation.
\subsection{The three-dimensional Janus metric}
\subsubsection{Time-dependent Janus deformation of BTZ metric}
The metric of the Janus black hole with its horizon radius $Lr_0$ is given by 
\begin{align} \label{eq:metric:tau}
  ds^2=L^2\,\frac{d\mu^2-d\tau^2+r_{0}^2\cos^2\tau d\theta^2}{g(\mu)^2},
\end{align}
where the only dimensionful quantity is the AdS radius $L$. The conformal factor $g(\mu)$ is defined as 
\begin{align}
  g(\mu)
  =\frac{\cn(\kappa_+\mu,k^2)}{\kappa_+\dn(\kappa_+\mu,k^2)}&&
  \kappa_\pm:=\sqrt{\frac{1\pm\sqrt{1-2\gamma^2}}{2}} &&
  k:=\frac{\kappa_-}{\kappa_+}.
\end{align}

\vspace{0.3 cm}

This metric is a one-parameter generalization of the BTZ black hole metric 
by ``Janus deformation parameter'' $0\le\gamma<1/\sqrt{2}$.
When $\gamma=0$, the factor $g(\mu)$ becomes $\cos\mu$ and then
the metric reduces to the BTZ metric, 
with its inverse temperature
\begin{align}\label{eq:beta}
  \beta = \f{2\pi}{r_0}\,,
\end{align}
in the unit of the AdS radius $L$.
The conformal boundaries $g(\mu)=0$ are located at $\mu=\pm \mu_{0}$, where $\mu_0:= K(k^2)/\kappa_+$ and $K(k^2)$ is the complete elliptic integral of the 1st kind $K(k^2):=\int_0^{\pi/2}d\theta/\sqrt{1-k^2\sin^2\theta}$.

\vspace{0.3 cm}

\subsubsection{Dual CFT coordinate $(t,\theta)$ and UV cutoff $\epsilon_\mathrm{CFT}$}
In applying AdS/CFT techniques, 
another time coordinate $\tanh r_0t:=\sin\tau$ is useful, because
the flat metric $-dt^2+d\theta^2$ of the dual CFT becomes manifest:
\begin{align} \label{eq:metric:t}
  ds^2
&= L^{2}\left[dy^2+\frac{r_0^2}{\ti{g}(y)^2\cosh^2r_{0}t}(-dt^2+d\theta^2)\right].
\end{align}
Here we have also replaced the radial coordinate $\mu$ with another one $y$ such that $\tanh y=\sn(\kappa_+\mu,k^2)$, measuring the proper length $dy=d\mu/g(\mu)$, and we have rewritten the factor $g(\mu)$ as
\begin{align}
  \ti{g}(y) 
:=g(\mu(y))
= \f{1}{\kappa_+\s{(1-k^2)\cosh^2{y}+k^2}}
= \s{\f{2}{1+\sqrt{1-2\gamma^2}\cosh 2y}}\,.
\end{align}
In this coordinate $y$, the origin $\mu=0$ corresponds to $y=0$ and the conformal boundaries $\mu=\pm\mu_0$ are located at $y\to \pm \infty$.

Near the  conformal boundaries  
$y\to \pm\infty$, the metric \eqref{eq:metric:t} approaches to the pure AdS metric in Poincar\'{e} coordinate
\begin{align}
  ds^2=L^{2}\frac{dz^2-dt^2+d\theta^2+\oo(z)}{z^2}
\end{align}
with the following identification
\begin{align}
  z:=\frac{2}{\sqrt[4]{1-2\gamma^2}\,r_0}e^{-|y|}\cosh r_0 t\,.
\end{align}
Hence
the CFT UV cutoff $\epsilon_\mathrm{CFT}$ is given as 
\begin{align}
 \epsilon_\mathrm{CFT}
  &=\frac{2}{\sqrt[4]{1-2\gamma^2}\,r_0}e^{-y_\infty}\cosh r_0 t_\infty, \label{eq:cutoff}
\end{align}
where 
$y_\infty(\gg1)$ is a bulk volume regulator and $t_\infty$ is the time $t$ in the CFT at $y=\pm y_\infty$.

\subsubsection{As a solution of Einstein-scalar theory}
\label{sec:Einstein-scalar}
This geometry is a solution of the three-dimensional Einstein-scalar system
\begin{align}
S=\frac{1}{16\pi G}\int d^3 x \sqrt{g} \left( R-g^{ab}\partial_{a}\phi\partial_{b}\phi+\frac{2}{L^2} \right),
\end{align}
with a scalar field configuration
\begin{align}
\phi
&=\phi_{0}+\sqrt{2}\l( \tanh^{-1}(k\sn(\kappa_+\mu,k^2))+\log\sqrt{1-k^2} \r) \nonumber\\
&=\phi_{0}+\sqrt{2}\l( \tanh^{-1}(k\tanh{y})+\log\sqrt{1-k^2} \r).
\end{align}
Note that the scalar field value $\phi_{+}:=\phi(y=\infty)$ on
the right boundary is different from the one $\phi_{-}:=\phi(y=-\infty)$ on the left boundary by
\begin{align}\label{eq:difference}
\phi_{+}-\phi_{-}=2\sqrt{2}\tanh^{-1}k
=\sqrt{2} \tanh^{-1} \sqrt{2} \gamma\,.
\end{align}

This three-dimensional system can be embedded in type IIB supergravity in ten dimensions
 with an appropriate ansatz \cite{Bak:2007jm}. 
Then in the same way as the standard D1-D5 black hole \cite{Strominger:1996sh,Callan:1996dv,Maldacena:1997re},
the boundary CFTs are given by the IR fixed points 
of the two-dimensional $\mathcal{N}=(4,4)$ supersymmetric $\SU(N_1)\times\SU(N_5)$ quiver field theories,
which turn out to be $\sigma$-models 
on the instanton moduli space $\mathcal{M}=M_4^{N_1N_5}/S_{N_1N_5}$.
The bulk scalar field $\phi$ is identified with the dilaton, 
and hence the boundary values $\phi_{\pm}$ are related to
the coupling constants $g_\pm$ of those boundary quiver theories \cite{Bak:2007jm}. 
In terms of the IR $\sigma$-models, 
this difference in the boundary values leads to the difference in the overall coefficients of the actions
on the two boundaries.

Although the difference between $\phi_+$ and $\phi_-$ \eqref{eq:difference} becomes
very large when we take $\gamma$ very close to $\sfrac{1}{\s{2}}$,
we can also take $\phi_0$  negatively large 
so that classical gravity does not break down.
In terms of the dual boundary theory, it requires that the theory is weakly coupled
in the sense of the Yang-Mills couplings
whereas it is strongly coupled in the viewpoint of the 't Hooft couplings,
as usual.

\subsection{Main differences from BTZ black hole}
\subsubsection{Causal shadow region}
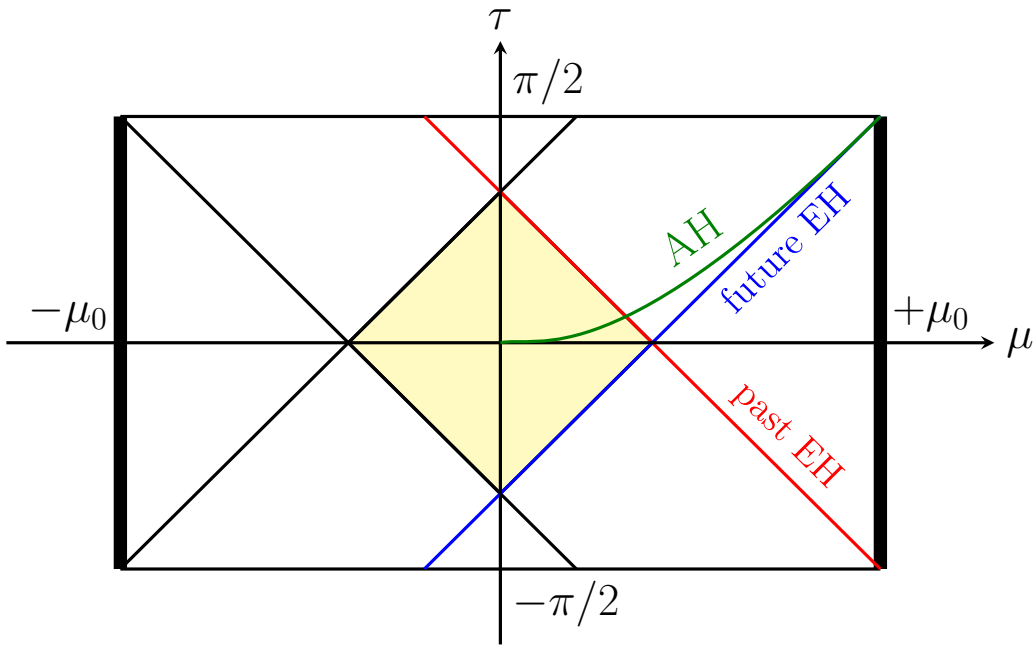
\begin{figure}[htb]
\begin{center}
\begin{tikzpicture}[>=stealth,very thick]
\draw[fill=yellow!30] (2,0) -- (0,2) -- (-2,0) -- (0,-2) -- cycle; 
\draw[line width=5pt] (5,3) --
(5,-3);
\draw[line width=5pt] (-5,3) -- (-5,-3);

\draw (-5,3) --
(5,3); 
\draw (-5,-3) --
(5,-3); 
\draw(-5,-3) --
(1,3); 
\draw(-5,3) --
(1,-3); 
\draw[red](5,-3) --
(2,0) node[midway,above,sloped] {{\large past EH}}-- (-1,3); 
\draw[blue](5,3) --
(2,0) node[midway,below,sloped] {{\large future EH}} --(-1,-3);
\draw[->] (-6.5,0) -- (6.5,0) node[right] {{\Large$\mu$}};
\draw[->] (0,-4) -- (0,4) node[above] {{\Large$\tau$}};
\draw[green!50!black] (0,0) .. controls (0.5,0.1)  and (1.5,-0.5) .. (5,3); 
\node[rotate=35,green!50!black] at (2.5,1.4) {{\Large AH}};

\draw (-5,0) node[above left]{{\Large$-\mu_0$}};
\draw (5,0) node[above right]{{\Large$+\mu_0$}};
\draw (0,3) node[above right]{{\Large$\pi/2$}};
\draw (0,-3) node[below right]{{\Large$-\pi/2$}};
\end{tikzpicture}
\end{center}
\caption{Penrose diagram of the three-dimensional time-dependent Janus black hole. 
The two conformal boundaries are located at $\mu=\pm\mu_0$ (thick lines), 
and the diagram is a wide rectangle because $\mu_0 \ge\pi/2$. 
The blue and red lines represent, respectively, the future and past event horizons which intersect with the right hand side boundary. The yellow shaded region corresponds to the ``causal shadow'' region, which is causally disconnected from the both boundaries.
The apparent horizons (green line) in time slices $\tau=\mathrm{const}.$ are located inside the future event horizon.}
\label{fig:penrose}
\end{figure}

By using  the conformally flat $(\mu,\tau)$ coordinate \eqref{eq:metric:tau}, one can draw the Penrose diagram of 
the time-dependent Janus black hole
geometry (see \fgref{fig:penrose}). 
The diagram is horizontally longer 
than that of the BTZ geometry, 
because the width $2\mu_0 = 2 K(k^2)/\kappa_+$
in the $\mu$ coordinate 
between the two conformal boundaries monotonically increases 
with the deformation parameter $\gamma$.

As a consequence, unlike the BTZ geometry ($\gamma=0$), 
the three-dimensional Janus black hole geometry 
($\gamma>0$) has a finite region causally disconnected 
from the both conformal boundaries $\mu=\pm\mu_0$.
Such regions are sometimes called ``causal shadow'' \cite{Headrick:2014cta,Fischetti:2014uxa}. 
It is an interesting question how the dual CFT encodes 
information on causal shadow regions. 
As a first step to answer this question, 
we will compute  holographic entanglement entropies 
in the Janus black hole geometry
in the next two sections, 
because holographic entanglement entropies 
can be affected by the inside of the causal shadow.

\subsubsection{Time-dependence}
Unlike the BTZ metric ($\gamma=0$), the Janus metric ($\gamma>0$) is time-dependent, that is to say, 
has no timelike Killing vector. As a result, its apparent horizon \begin{align}
  \tan\tau=-\frac{d}{d\mu}\log g(\mu)
\end{align}
in a time slice $\tau=\mathrm{const}.$ becomes different from the event horizon $\tau-\pi/2=\mu-\mu_0$ (See \fgref{fig:penrose}).

\subsection{The CFT interpretation of the Janus black hole}
When $ \gamma = 0$, 
the Janus black hole reduces to the ordinary eternal BTZ black hole,
which  is dual to the thermofield double state
\cite{Maldacena:2001kr,Balasubramanian:1998de}
\begin{align}
  |\Psi \rangle 
  =\frac{1}{\sqrt{Z}} \sum_{n} e^{-\frac{\beta}{2}E_{n}}  
  |E_{n} \rangle 
  |E_{n} \rangle\,.
\end{align}
The inverse temperature $\beta$ is given by \eqref{eq:beta}.

If we turn on the parameter $\gamma$, the Hamiltonian $H_+$ on the right boundary and $H_-$ on the left boundary become different,
as was explained in \S\ref{sec:Einstein-scalar}.
Hence it is natural to conjecture \cite{Bak:2007qw} that
the Janus black hole is dual to a state 
\begin{align}\label{eq:cftstate}
  |\Psi \rangle 
  =\frac{1}{\sqrt{Z}} \sum_{(m,n)} 
  e^{-\frac{\beta}{4}(E_{n}^-+E_{m}^+)}
  \langle E^{+}_{m}| E^{-}_{n} \rangle 
  |E^{+}_{m} \rangle 
  |E^{-}_{n} \rangle\,.
\end{align}
This conjecture has passed some nontrivial checks. 
For example, the one point function of the Lagrangian density was computed 
both on gravity and CFT sides, which agrees 
up to the second order in $\gamma$ \cite{Bak:2007qw}.

\section{Calculation of Holographic Entanglement Entropy} 
\label{sec:timeevolution}
In this section, we compute a holographic entanglement entropy 
on the three-dimensional Janus black hole geometry
to study an entanglement between the left and right CFT's. 
We take our subsystem $A$ to be 
two disjoint same intervals $-\theta_{\infty}\leq \theta\leq  \theta_{\infty}$ in each of the left and right CFT at a fixed time $t= t_{\infty}$ (see \fgref{fig:phases}). 

\subsection{Covariant holographic entanglement entropy}

It has been conjectured \cite{Hubeny:2007xt} that for a given bulk geometry,
the entanglement entropy of the  dual CFT state is given by the area of the extremal surface\footnote{
The original holographic entanglement entropy formula \cite{Ryu:2006bv,Ryu:2006ef} 
(later proven in \cite{Lewkowycz:2013nqa})
with minimal surface prescription is only applicable to static bulk geometries.
The extension \eqref{eq:covSA} to general geometries is achieved by
just replacing the ``minimum'' on the time-slice by the ``extremum'' in the 
spacetime.

There are many equivalent constructions which look different.
The extremal surface explained above (called $\mathcal{W}$ in the original paper \cite{Hubeny:2007xt}),
the surface with vanishing traces of extrinsic curvatures ($\mathcal{Y}_{ext}$),
and 
a surface constructed by using light-sheets ($\mathcal{Y}_{\mathcal{A}_t}^{min}$),
are eventually all equivalent.
See the original paper for the detail.
} 
in the bulk which are anchored to $\partial A$ in the conformal boundary, 
\begin{align}\label{eq:covSA}
S_{A}= \ext\frac{A(\gamma_{A})}{4G_N},
\end{align}
where $G_N$ is the three-dimensional Newtonian constant.
The extrema is chosen among the surfaces  $\gamma_{A}$ 
which are homologous to the subsystem $A$ 
and satisfying $\partial A= \partial \gamma_{A}$. 
If there are multiple extremal surfaces, 
we should choose the one with the minimum area among them.

\begin{figure}[htbp]
\centering
\begin{tikzpicture}[thick,>=stealth]
   \draw (-2,-1)--(0,0.5)--(0,4.5)--(-2,3)
   node[midway,above,sloped]{} --(-2,-1);
   \draw[->] (-2-0.4,1-0.3)--(0+0.4,2.5+0.3) node[right] {$\theta$};
   \draw[->] (-1,-0.6)--(-1,4.2) node[left] {$t$};
   \draw[red](-1.5,11/8)--(-0.5,17/8);
   \draw[] (-1,2) node[above,left,red] {{\Large$A$}};
   \draw[blue!50!cyan] (-1.5,11/8) arc (-90:60:2 and 0.4) ;

   \draw (2,-1)--(4,0.5)--(4,4.5)--(2,3)
   node[midway,above,sloped]{} --(2,-1);
   \draw[->] (2-0.4,1-0.3)--(4+0.4,2.5+0.3) node[right] {$\theta$};
   \draw[->] (3,-0.6)--(3,4.2) node[left] {$t$};
   \draw[red](2.5,11/8)--(3.5,17/8);
   \draw[] (4,1.8) node[above,left,red] {{\Large$A$}};
   \draw[blue!50!cyan] (2.5,11/8) arc (-120:-270:2 and 0.4) ;

   \draw[] (1,1) node[above,left,blue] {{\Large$\gamma_A$}};
   \draw[->](-1,1.7)--(1,1.7) node[right] {$y$};

   \draw (-1.5,-2) node[right] {($a$) disconnected phase};
\end{tikzpicture}
\hspace{10pt}
\begin{tikzpicture}[thick,>=stealth]
   \draw (-2,-1)--(0,0.5)--(0,4.5)--(-2,3)
   node[midway,above,sloped]{} --(-2,-1);
   \draw[->] (-2-0.4,1-0.3)--(0+0.4,2.5+0.3) node[right] {$\theta$};
   \draw[->] (-1,-0.6)--(-1,4.2) node[left] {$t$};
   \draw[red](-1.5,11/8)--(-0.5,17/8);
   \draw[] (-1,2) node[above,left,red] {{\Large$A$}};
   \draw[blue!50!cyan] (-1.5,11/8) -- (2.5,11/8) ;
   \draw[blue!50!cyan] (-0.5,17/8)--(3.5,17/8);

   \draw (2,-1)--(4,0.5)--(4,4.5)--(2,3)
   node[midway,above,sloped]{} --(2,-1);
   \draw[->] (2-0.4,1-0.3)--(4+0.4,2.5+0.3) node[right] {$\theta$};
   \draw[->] (3,-0.6)--(3,4.2) node[left] {$t$};
   \draw[red](2.5,11/8)--(3.5,17/8);
   \draw[] (4,1.8) node[above,left,red] {{\Large$A$}};

   \draw[->](-1,1.7)--(1,1.7) node[right] {$y$};
   \draw[] (1.3,1) node[above,left,blue] {{\Large$\gamma_A$}};

   \draw (-1.5,-2) node[right] {($b$) connected phase};
\end{tikzpicture}
\caption{The subsystem $A$ (two red lines) is taken as two disjoint intervals of the same length $\Delta\theta=2\theta_\infty$ in the right and left boundary (two black squares). The extremal surface $\gamma_A$ (blue lines) has two phases: disconnected phase ($a$) and connected phase ($b$).}
\label{fig:phases}
\end{figure}
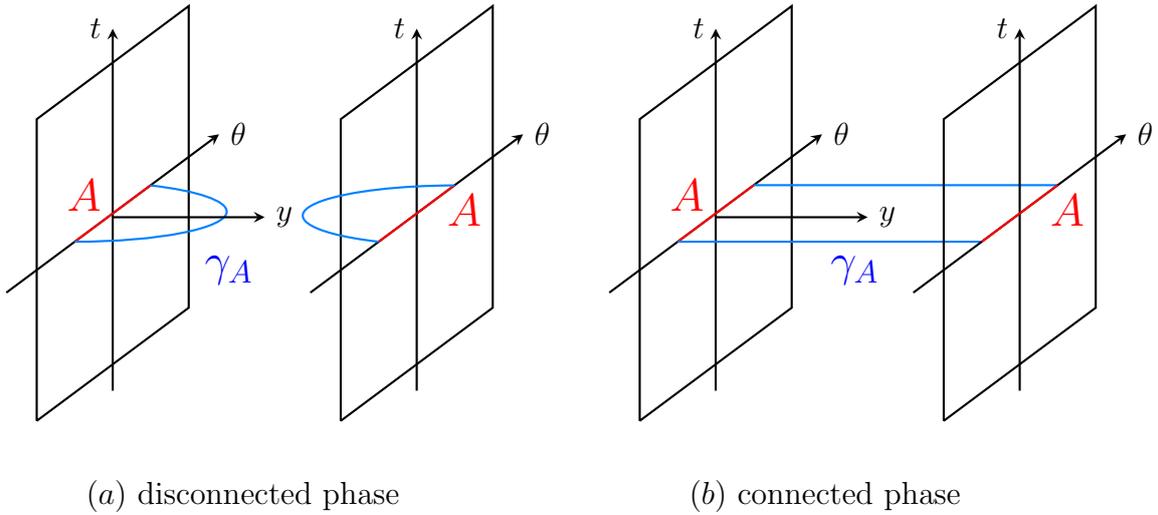

In the current setup with the subsystem $A=\{(\pm y_\infty,t_\infty,\theta);-\theta_\infty\le \theta\le\theta_\infty\}$ in the Janus black hole geometry \eqref{eq:metric:t},
the extremal surface can take two types of topologies (see \fgref{fig:phases}), ``connected phase''  and ``disconnected phase'', 
like the usual BTZ black holes \cite{Hartman:2013qma}.
The disconnected type consists of two geodesics which start from and end at the same boundary (see \fgref{fig:phases} ($a$)); 
starting from $(\pm y_\infty,t_\infty,-\theta_\infty)$, 
turning around at $(\pm y_*,t_*,0)$ and ending at $(\pm y_\infty,t_\infty,\theta_\infty)$.
The connected type consists of two geodesics which connect the two boundaries (see \fgref{fig:phases} ($b$));
starting from $(y_\infty,t_\infty,\pm\theta_\infty)$ and ending at $(-y_\infty,t_\infty,\pm\theta_\infty)$.

In the following, we will obtain and solve differential equations for each type of extremal surfaces.
Identifying the area functional
\begin{align}  \label{eq:eom:area}
  A[t(y),\theta(y)]
  &=L\int\!dy\,\sqrt{1+\frac{r_0^2}{\ti{g}(y)^2\cosh^2r_0t}(-\dot{t}^2+\dot{\theta}^2)}
\end{align}
with a classical action for dynamical variables $t(y)$ and $\theta(y)$ as for ``time'' $y$,
this problem reduces to just an Euler-Lagrange problem.
Here the dot (\,$\dot{}$\,) represents the ``time'' derivative $d/dy$.
We will see that the disconnected surface, 
as well as the connected one, 
can penetrate the event horizon, 
and both of their areas are dependent on the boundary time $t_{\infty}$.
The phase transition between these two types will be discussed in \secref{sec:phase_trans}. 

\subsection{Extremal areas in connected phase}
For connected surfaces, the area functional is extremized 
when $\theta=\mathrm{const.}$ ($=\pm\theta_\infty$).
Then the ``action'' \eqref{eq:eom:area} becomes
\begin{align}
  A[t(y)]/L
  =\int_{-y_\infty}^{y_\infty} dy\sqrt{1-\f{r_0^2\,\dot{t}^2}{\ti{g}(y)^2\cosh^2{r_0t}}}\,,
\end{align}
for each of the two pieces of the surface ($\theta=\pm\theta_\infty$).
Here $y_{\infty}$ is the bulk volume regulator, which also regulates the area functional. 
This functional  has one conserved charge $E$:
\begin{align}
  E:=\frac{\delta A/L}{\delta \dot{t}}
  &=\frac{-r_0^2\,\dot{t}}{\ti{g}(y)\cosh{r_0t}\sqrt{\ti{g}(y)^2\cosh^2{r_0t}-r_0^2\,\dot{t}^2}}
  \nonumber\\
  &\Leftrightarrow 
  \quad
  \dot{t}=\f{-E\,\ti{g}(y)^2\cosh^2{r_0t}}{\s{r_0^2+E^2\,\ti{g}(y)^2\cosh^2{r_0t}}}\,,
\end{align}
associated to its $t$-translation symmetry. 
But this charge $E$ vanishes, because $\dot{t}$ cannot change its sign 
and we have the boundary condition $\int_{-y_\infty}^{y_\infty} \dot{t}~dy =t_\infty-t_\infty=0$.
In the result, the total area of the connected extremal surface can be explicitly calculated as
\begin{align}
  A_c(t_\infty,\theta_\infty)/L&=2\times2y_\infty
\nonumber\\
  &=4\log\frac{2\cosh r_{0}t_\infty}{r_{0}\epsilon_\mathrm{CFT}}
  -\log(1-2\gamma^2).
\label{eq:Ac}
\end{align}
To derive this, we used the relation between the regulator $y_{\infty}$ and the CFT cutoff $\epsilon_\mathrm{CFT}$ \eqref{eq:cutoff}. 
For later purposes, it is convenient to define the notion of 
``renormalized'' area which is given by 
\begin{align}
  A_c^{\mathit{(ren)}}/L &\equiv A_c( t_\infty, \theta_{\infty})/L+4\log \epsilon_\mathrm{CFT} \nonumber \\ 
  &=4\log\frac{2\cosh r_{0}t_\infty}{r_{0}}
  -\log(1-2\gamma^2)\,.
\label{eq:Ac:ren}
\end{align}
Note that the connected surface area becomes arbitrarily large in $\gamma^2 \rightarrow \frac{1}{2}$ limit. 
This illuminates the fact that
the length of the wormhole behind the Janus black hole becomes infinitely long in this limit.

\subsection{How to calculate extremal areas in disconnected phase}
In this subsection, we represent the area of the disconnected surfaces 
as a function of boundary coordinates ($t_\infty, \theta_\infty$). 
The disconnected surfaces consist of two disjoint geodesics, 
one of which is located in the right region $y>0$ and the other is in the left region $y<0$. 
In what follows, we take the $y>0$ part of the surfaces,
because the $y<0$ part can be identified with $y>0$ part 
by the parity transformation $y \to -y$. 

The disconnected surface area $A_{dc}$ can be given by an integral
from the boundary $y=y_\infty$ to the returning point $y=y_*$,
at which the derivative $\dot{\theta}$ of the surface $(t,\theta)=(t(y),\theta(y))$ diverges.
Its location ($y_*,t_*$) is determined by the boundary coordinates $(t_\infty, \theta_\infty)$,
and so we can represent the area $A_{dc}$ as a function of $( t_{\infty}, \theta_{\infty} )$
by substituting the expression $y_*=y_* (t_{\infty},\theta_{\infty})$ into the area integral.
\subsubsection{Solving the equation of motion} \label{subsubsection:basic ingredients}
We can solve the Euler-Lagrange equations for $t(y)$ and $\theta(y)$ in the following way. 

The action \eqref{eq:eom:area} has one conserved charge $J$,
\begin{align}
  J:&=\frac{\delta A/L}{\delta \dot{\theta}}
  =\frac{1}{\ti{g}(y)\cosh r_{0} t}\frac{r_{0}^2\dot{\theta}}{\sqrt{\ti{g}(y)^2\cosh^2 r_{0} t+ r_{0}^2(-\dot{t}^2+\dot{\theta}^2)}}
  \label{eq:eom:chargeJ}
\end{align}
associated to its $\theta$ translation symmetry.
This charge $J$ can be expressed by the returning point location $(y_*,t_*)$ as $J=r_0(\ti{g}(y_*)\cosh r_0t_*)^{-1}$, 
because $\dot{\theta}$ in (\ref{eq:eom:chargeJ}) diverges at the returning point. 
With the aid of this constant charge $J$, 
the equation of motion for $t(y)$ can be rewritten into an equation for $t(\theta)$ without any $\ti{g}(y)$ dependence:
\begin{align}
  \frac{d}{dy}\frac{\delta A}{\delta \dot{t}}-\frac{\delta A}{\delta t}=0
  &\Leftrightarrow
  \frac{d}{dy}\l( J \frac{\dot{t}}{\dot{\theta}}\r)
   =J r_{0}\frac{-\dot{t}^2+\dot{\theta}^2}{\dot{\theta}}\tanh r_0t \\
  &\Leftrightarrow
  \frac{d^2t}{d\theta^2}
   =r_{0}\left[1-\left(\frac{dt}{d\theta}\right)^2\right]\tanh r_{0} t,
\end{align}
whose general solution is given by
$\sinh{r_{0}t}=\sinh{A}\cosh{r_{0}(\theta+B)}$
with some constants $A$, $B$. 
These constants $A$, $B$ are determined by geometrical conditions $\theta|_{y=y_*}=0$ and $dt/d\theta|_{y=y_*}=0$ as 
\begin{align}
  \sinh r_{0} t=\sinh r_{0} t_*  \cosh r_{0} \theta.
    \label{eq:eom:OBK}
\end{align}
This relation allows us to erase $\theta$ in
 \eqref{eq:eom:chargeJ},
yielding a 1st order differential equation of $t$:
\begin{align}
  \dot{t}
  &=\frac{\cosh r_{0} t}{r_0\ti{g}(y_*)\cosh r_{0}t_*}\sqrt{
   \frac{\cosh^2 r_{0}t-\cosh^2 r_{0} t_*}{1-(\ti{g}(y)/\ti{g}(y_*))^2}},
   \label{eq:eom:tdot}
\end{align}
which has a unique solution
\begin{align}
  \sqrt{1-\frac{\sinh^2 r_{0}t_*}{\sinh^2 r_{0}t}}
  ~&(=\tanh\theta)=\cosh r_{0} t_*\tanh\l[\int_{y_*}^y dy~\frac{\ti{g}(y)^2}{\sqrt{\ti{g}(y_*)^2-\ti{g}(y)^2}}\r]\,,
  \label{eq:eom:solution}
\end{align}
with an initial condition $t(y_*)=t_*$. 
This expression gives the unique solution $(t(y),\theta(y))$ of the equations of motion, 
in terms of the returning point location $(y_*,t_*)$. 

\subsubsection{Returning point ($y_*,t_*$)}
The boundary condition $ (t(y_\infty)=t_\infty, \; \theta (y_{\infty})= \pm \theta_\infty$)
determines  $t_*$ by \eqref{eq:eom:OBK}  as
\begin{align}
  \sinh r_{0}t_*=\frac{\sinh r_{0} t_\infty}{\cosh r_{0} \theta_\infty},
  \label{eq:eom:tstar}
\end{align}
 and  $y_*$  by \eqref{eq:eom:OBK} and \eqref{eq:eom:solution} as
\begin{align}
  \sinh\left[\int_{y_*}^{y_\infty}dy\, \frac{\ti{g}(y)^2}{\sqrt{\ti{g}(y_*)^2-\ti{g}(y)^2}}\right]
  =\frac{\sinh r_{0}\theta_\infty}{\cosh r_{0}t_\infty}\,.
  \label{eq:eom:ystar}
\end{align}

\begin{figure}[htbp]
  \begin{center}
    \includegraphics[width=8cm]{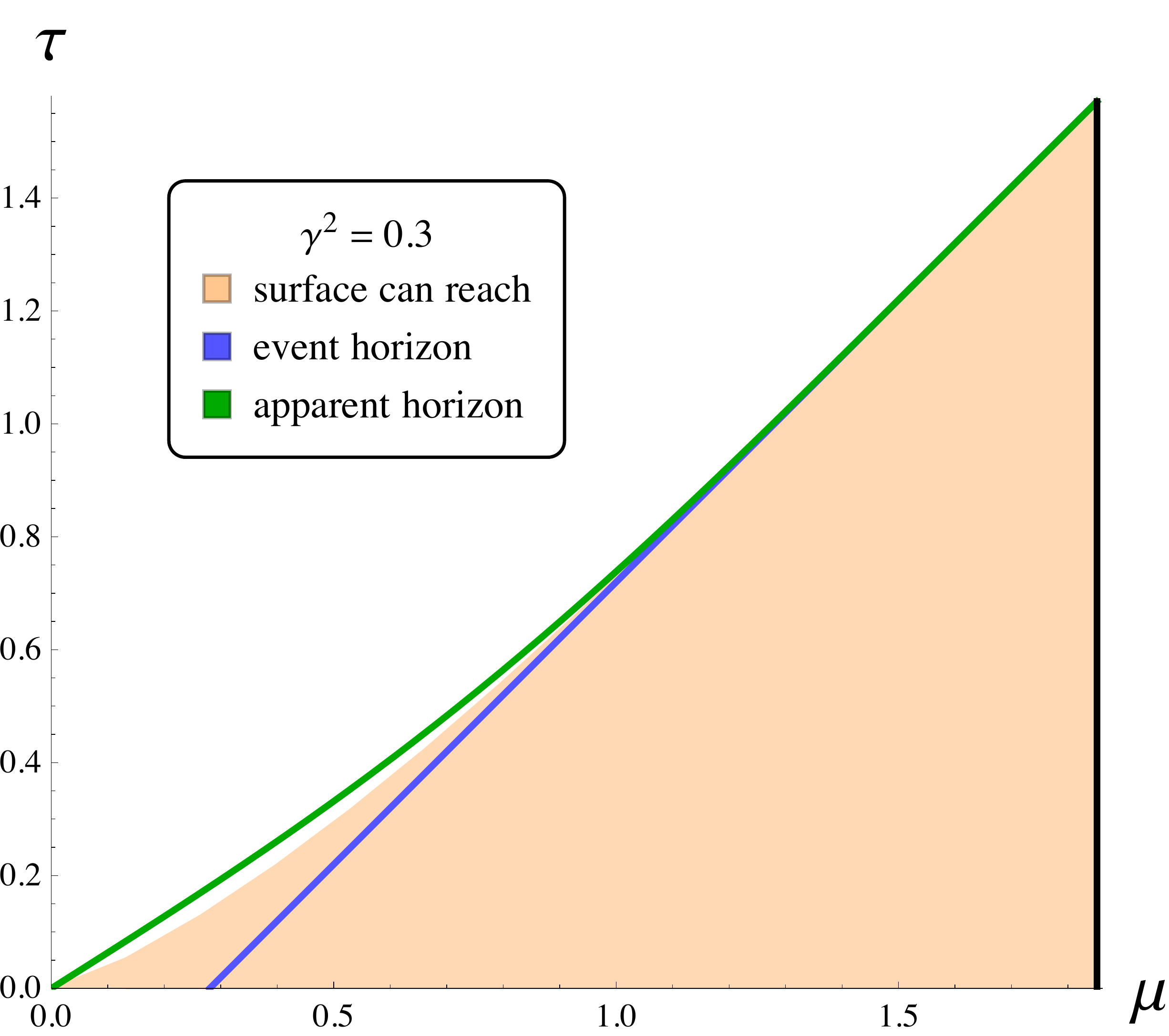}
  \end{center}
  \caption{How deeply the extremal surfaces in the disconnected phase can go inside the Janus black hole (with $\gamma^2=0.3$ in the figure). The shaded orange region represents where the extremal surfaces can pass through. The extremal surfaces can go beyond the event horizon (blue line), but cannot go beyond the apparent horizon (green line). }
  \label{fig:SurfaceBarrier}
\end{figure}

Note that there are bulk points which cannot be returning points
for any boundary value $(t_\infty,\theta_\infty)$,
and that there exists a region which cannot be reached by the connected surfaces (see Figure \ref{fig:SurfaceBarrier}).
It might be interesting that the surface can go beyond the event horizon but cannot go beyond the apparent horizon.

\subsubsection{Extremal surface area }
Plugging  \eqref{eq:eom:OBK} and \eqref{eq:eom:tdot}  into the definition of the surface area \eqref{eq:eom:area},
we obtain the disconnected extremal surface area
\begin{align}
    A_{\mathit{dc}}(t_\infty,\theta_\infty)/L
  &=4\int_{y_*}^{y_\infty} dy~ \frac{\ti{g}(y_*)}{\sqrt{\ti{g}(y_*)^2-\ti{g}(y)^2}}\,,
  \label{eq:Adc:y}
\end{align}
as a function of  $(t_\infty,\theta_\infty)$,
with $y_*$ implicitly determined by $(t_\infty,\theta_\infty)$ through \eqref{eq:eom:ystar}.

Note that this area has a UV divergence $-4\log\epsilon_\mathrm{CFT}$, 
because  $\ti{g}(y) \to 0$ at the each boundary and
\begin{align}
  A_{\mathit{dc}}/L
  \to 4\int^{y_\infty}\!dy
  \sim 4y_\infty
  = 4\log\frac{2\cosh r_{0}t_\infty}{\s[4]{1-2\gamma^2\,}\, r_{0}\epsilon_\mathrm{CFT}}\,.
\end{align}
This UV divergence can be renormalized as
\begin{align}
  A_{\mathit{dc}}^\mathrm{(ren)}/L& \equiv A_{\mathit{dc}}/L+4\log\epsilon_\mathrm{CFT} \nonumber\\
  &= A_{\mathit{dc}}/L -\log(1-2\gamma^2) + 4\log\f{2\cosh{r_0t_\infty}}{r_0} -4y_\infty\,.
  \label{eq:area:ren}
\end{align}

\subsection{Some limits of  extremal surface areas in disconnected phase}
It is generally difficult to calculate the area of the disconnected surface. 
In this subsection, we address some limits in which this area is explicitly calculable. 
In Sections~\ref{subsubsec:early} and \ref{subsubsec:late}, 
we  compute the disconnected surface area $A_{\mathit{dc}}^\mathrm{(ren)}(t_\infty,\theta_\infty)$ 
with a large subsystem ($\theta_\infty \gg r_0^{-1}$), 
in the early time ($t_\infty\ll\theta_\infty$) 
and late time ($t_\infty\gg\theta_\infty$) limit. 
In \secref{subsubsec:gamma}, we compute the area in the small $\gamma$ limit.

\subsubsection{Early time limit for large subsystem ($\theta_\infty \gg t_\infty$)}
\label{subsubsec:early}
In this parameter region, the returning point $(y_*,t_*)$ is close to the origin $(0,0)$, which can be seen as follows.
The $t_*$ is determined by \eqref{eq:eom:tstar} as
\begin{align}
  r_0t_*\simeq2e^{-r_0\theta_\infty}\sinh r_0t_\infty 
\quad (\ll 1)\,,
  \label{eq:tstarsmall}
\end{align}
where we used\, $\cosh r_{0} \theta_\infty \simeq e^{r_0 \theta_\infty}/2$\,  and\, $\sinh r_0 t_*\simeq r_0t_*$.
The $y_*$ is determined by \eqref{eq:eom:ystar} as
\begin{align}
\label{eq:ystarsmall}
  r_{0}\theta_\infty-\log\cosh r_{0} t_\infty\simeq\int_{y_{*}}^{y_\infty}dy~ \frac{\ti{g}(y)^2}{\sqrt{\ti{g}(y_*)^2-\ti{g}(y)^2}}
\quad (\gg 1)\,.
\end{align}
This means $y_*\ll1$,
because the left hand side of  (\ref{eq:ystarsmall}) is large 
while the integral of  the right hand side is a monotonically decreasing function of $y_*$, diverging at $y_*\to0$. 
In fact, the right hand side integral can be evaluated as
\begin{align}
  r_{0} \theta_\infty-\log\cosh r_{0} t_\infty
  \simeq  -\frac{1}{\sqrt{\kappa_+^2-\kappa_-^2}}\log\left[\frac{\kappa_++\sqrt{\kappa_+^2-\kappa_-^2}}{4} y_*\right]\,
  \label{eq:early:mustar}
\end{align}
in the limit $y_*\to0$, by changing the integration variable from $y$ to $z:=\tanh y$.

By solving this for $y_{*}$ and plugging it into 
 \eqref{eq:area:ren}, 
we obtain the renormalized area $A_{\mathit{dc}}^\mathrm{(ren)}(t_\infty,\theta_\infty)$ as a function of $(t_\infty,\theta_\infty)$.
This can be carried out by evaluating
the integration of \eqref{eq:Adc:y} similarly as
\begin{align}
  &\int_{y_*}^{y_\infty} dy~ \frac{\ti{g}(y_*)}{\sqrt{\ti{g}(y_*)^2-\ti{g}(y)^2}}
\nonumber\\
  &\simeq
  -\frac{\kappa_+}{\sqrt{\kappa_+^2-\kappa_-^2}}\log\left[\frac{\kappa_++\sqrt{\kappa_+^2-\kappa_-^2}}{4}y_*\right]
  -\log\left[\frac{\kappa_++\sqrt{\kappa_+^2-\kappa_-^2}}{\sqrt{\kappa_+^2-\kappa_-^2}}\right]+y_\infty\,
\end{align}
in the limit $ y_*\to0$.
We can delete $y_*$ by \eqref{eq:early:mustar}, which results in
\begin{align}
\label{eq:earlyarea}
    A_{\mathit{dc}}^\mathrm{(ren)}(t_\infty,\theta_\infty)/L
  \simeq 4\kappa_+r_0\theta_\infty+4(1-\kappa_+)\log\cosh r_0 t_\infty-4\log\left[\frac{\kappa_++\sqrt{\kappa_+^2-\kappa_-^2}}{2}r_0\right].
\end{align}
Note that the area linearly grows with both $t_\infty$ and $\theta_\infty$
with different coefficients,
when $\theta_\infty\gg t_\infty\gg r_0^{-1}$.

\subsubsection{Late time limit for large subsystem ($t_\infty\gg\theta_\infty\gg r_0^{-1}$)}

\label{subsubsec:late}
In this parameter region, 
\eqref{eq:eom:tstar} and \eqref{eq:eom:ystar}
lead to
\begin{align}\label{eq:latetime:approx}
  2e^{-r_0 t_*}
  \simeq
  \int_{y_*}^{y_\infty}\!\!
  \frac{\ti{g}(y)^2dy}{\sqrt{\ti{g}(y_*)^2-\ti{g}(y)^2}}
  \simeq
  e^{r_{0}(\theta_\infty-t_\infty)}
  \quad (\ll1)\,,
\end{align}
where we used $\sinh{x}\simeq\cosh{x}\simeq e^x/2$ for $x\gg 1$
and $\sinh{x}\simeq x$ for $x\ll 1$.
This in turn implies $y_*\gg 1$, 
therefore the integrals in 
\eqref{eq:latetime:approx} and \eqref{eq:Adc:y}
can be respectively approximated as
\begin{align}
  \label{eq:latetime:integral:ystar}
  \int_{y_*}^{y_\infty}\!\!
  \frac{\ti{g}(y)^2dy}{\sqrt{\ti{g}(y_*)^2-\ti{g}(y)^2}}  
  &\simeq
  \f{2}{\s[4]{1-2\gamma^2\,}}e^{-y_*}\,,
\\
  \label{eq:latetime:integral:Adc}
  \int_{y_*}^{y_\infty}\!\!
  \frac{\ti{g}(y_*)dy}{\sqrt{\ti{g}(y_*)^2-\ti{g}(y)^2}}  
  &\simeq
  y_{\infty} - y_* + \log{2}\,,
\end{align}
where we also used $y_\infty-y_*\gg 1$.
By substituting
\eqref{eq:latetime:integral:Adc} into \eqref{eq:Adc:y},
and by erasing $y_\infty$ and $y_*$ with the aid of 
\eqref{eq:cutoff},
\eqref{eq:latetime:approx} and \eqref{eq:latetime:integral:ystar},
we can evaluate 
the renormalized area $A_{\mathit{dc}}^\mathrm{(ren)}$ \eqref{eq:area:ren} as
\begin{align}\label{eq:latearea}
  A_{\mathit{dc}}^\mathrm{(ren)}(t_\infty,\theta_\infty)/L
  \simeq 
  4(r_0\theta_\infty - \log{r_0})\,,
\end{align}
which does not depend on either of $\gamma$ or $t_\infty$.
Then in particular,
it coincides with the BTZ ($\gamma^2=0$) result.

\subsubsection{Up to the lowest order of $\gamma^2$}
\label{subsubsec:gamma}So far, we have seen the early time $ t_{\infty}\ll \theta_{\infty}$ and the late time $\theta_{\infty}\ll t_{\infty}$
behavior of the disconnected surface area for general $\gamma$,
but the surface phase transition discussed in the next section typically occurs at the intermediate time
region $t_{\infty} \sim \theta_{\infty}$. 
To obtain an analytic expression applicable to the whole time region,
let us evaluate the surface area up to the lowest order of the deformation $\gamma^2$.  
By expanding the relation (\ref{eq:eom:ystar}) and the area integral (\ref{eq:Adc:y}) up to the order of $\gamma^2$, 
we get
\begin{align}
  A_{\mathit{dc}}^{\mathit{(ren)}}/L
&=
  4\log\l(\f{2}{r_0}\sinh{r_0\theta}\r)
- \l(\f{3F^2+2}{2\s{1+F^2}}\coth^{-1}\l(\s{1+F^2}\r)-\f{3}{2}\r)\gamma^2 + \mathcal{O}(\gamma^4)\, ,
\label{eq:area:ren:expansion}
\end{align}
where 
\begin{align}
  F(t,\theta) = \f{\cosh{r_0t_\infty}}{\sinh{r_0\theta_\infty}}\,.
\end{align}
The detail of this calculation is explained in \appref{sec:gamma_expand}. 
Note that when $\gamma=0$, it reduces to the usual thermal result. 
In the early ($F\ll1$) and late ($F\gg1$) time limits, it respectively reproduces the results 
\eqref{eq:earlyarea} and \eqref{eq:latearea}
in the previous section.

\section{Time Evolution of Entanglement Entropy and Phase Transition} 
\label{sec:phase_trans}

Here we discuss the time-dependent behavior of the holographic entanglement entropy. 
Since there are two extremal surfaces in the bulk geometry, 
the holographic entanglement entropy $S_{A}$ is given by choosing the one with the minimum area among them, 
\begin{align}
  S_{A}= \frac{1}{4G_N} \; \min\left\{ A_c, A_{\mathit{dc}} \right\}, 
\end{align}
where $A_c$ and $A_{\mathit{dc}}$ 
are given by \eqref{eq:Ac} and \eqref{eq:Adc:y} respectively.
As we will see below, the entropy $S_A$ behaves very differently depending on 
the deformation parameter $\gamma$. 

\paragraph{$\bm{\gamma=0}$}
When $\gamma=0$, the spacetime reduces to the BTZ black hole, 
and does not contain any causal shadow region. 
The connected and disconnected surface areas are respectively given by 
\begin{align}
A_c(t_{\infty}, \theta_{\infty})/L=4 \log\l(\frac{2 \cosh r_{0}t_{\infty}}{r_{0} \epsilon_\mathrm{CFT}}\r)\,, 
\qquad 
A_{\mathit{dc}}(t_{\infty}, \theta_{\infty})/L= 4\log\l(\f{2 \sinh{r_0\theta_{\infty}}}{r_0 \epsilon_\mathrm{CFT}}\r)\,.
\end{align}
Let us take a sufficiently large subsystem 
$r_0\theta_{\infty}\gg 1$.
The entropy initially grows linearly with time because the connected surface is chosen in accordance with $A_c < A_{dc}$, and stops growing at a critical time $t_{\infty}=t_c\simeq \theta_{\infty}$. After the critical time, it ends up with a constant value, double the value of the thermal entropy, because the disconnected surface becomes chosen in accordance with $A_{dc}<A_c$.    

This time-dependent behavior such as the sharp phase transition can be also be observed on the CFT side, 
since the time-scale of the transition is given by $\beta$ \cite{Hartman:2013qma}
and now
$r_0\theta\gg 1$ implies $t_c \gg \beta$.
Furthermore, 
the initial entanglement entropy at $t_\infty=0$ can be identified with the contribution from the boundary of $A$ (4 points).
The time-dependent behavior can be intuitively understood in the so-called quasi-particle picture \cite{Calabrese:2005in}. 
In this picture, we assume that 
a pair creation of entangled quasi-particles occurs at every spatial point
at the initial time,
and that the pair propagate in opposite directions at the speed of light. A pair contributes to the entanglement
entropy if one of the pair is inside the subsystem and the 
other of the pair is outside the subsystem.
This picture correctly reproduces the linear growth and saturation of the entanglement entropy.

\paragraph{$\bm{0<\gamma^2 \ll \frac{1}{2}}$}
When $0<\gamma^2 \ll \frac{1}{2}$, 
the story is quite similar to the BTZ case, $\gamma=0$.
The entanglement entropy grows up until
a critical time $t_{\infty}=t_{c}\simeq\theta_{\infty}$, 
when the areas of the two surfaces become equal
and a phase transition takes place.
At that time, the growth rate of the entanglement entropy suddenly decreases discontinuously, but does not immediately become zero, unlike the BTZ case.
The entanglement entropy continues to grow very slowly and converges to a constant 
independent of $\gamma$. 
Hence the final value is identical with that of the BTZ case, $\gamma=0$, in particular. 

\vspace{0.1cm}

Another important difference 
from the BTZ black hole case
is that 
the initial entanglement entropy includes
an additional positive term $(-\f{L}{4G_N}\log(1-2\gamma^2))$.
This term can be regarded as a kind of boundary entropy, 
which is the contribution of defects in the system \cite{Affleck:1991tk} 
(see also \cite{Azeyanagi:2007qj} for the holographic realization). 
Note that in our system the defect is localized along the Euclidean time direction.

\begin{figure}[htbp]
  \begin{center}
    \includegraphics[width=12cm]{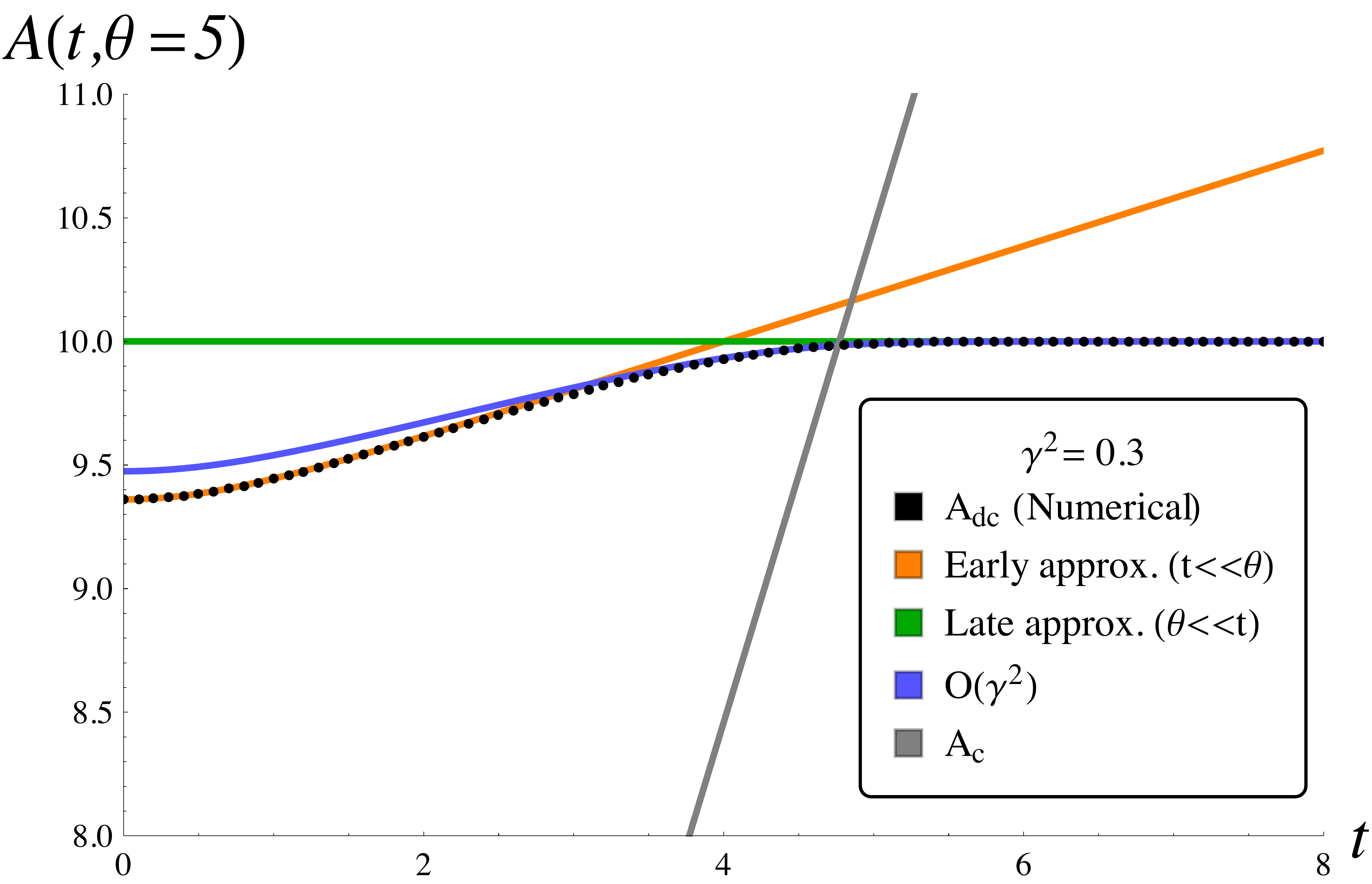}
  \end{center}
  \caption{The time $t$ dependence of the extremal surface area $A$ for a subsystem $\theta=5$, in the disconnected phase (black dotted line, numerically obtained) and in the connected phase (gray line).
The phase transition from the connected phase to the disconnected phase occurs at their intersection point $t=t_c$. The disconnected phase surface area $A_{\mathit{dc}}$  is initially well approximated by the early time limit approximation \eqref{eq:earlyarea} (orange line),
and finally well approximated by the late time limit approximation \eqref{eq:latearea} (green line). The whole time-dependence of $A_{\mathit{dc}}$ is qualitatively reproduced by the calculation \eqref{eq:area:ren:expansion} up to $\oo(\gamma^2)$  (blue line).}
  \label{fig:HEE_TimeDependence}
\end{figure}

It is difficult to determine the critical time $t_{c}$ analytically for arbitrary $\gamma$ and $\theta_{\infty}$, 
because one need to evaluate the disconnected surface area \eqref{eq:Adc:y} around the difficult time region $t_{\infty} \sim \theta_{\infty}$. 
Here we evaluate it perturbatively around $\gamma=0$ up to the second order. 
The detail of the calculation is given in \appref{sec:gamma_expand}.
By equating \eqref{eq:Ac} and \eqref{eq:area:ren:expansion}, we obtain
\begin{align}\label{eq:tc}
  t_c \simeq \theta_{\infty} - 2.058\gamma^2 + \mathcal{O}(\gamma^4)\,.
\end{align}
Note that the coefficient of $\gamma^2$ does not depend on the size of the subsystem $\theta_{\infty}$ or $r_{0}$. 

\vspace{0.1cm}

We can also solve the equations of motion for the disconnected extremal surface numerically,
to calculate the accurate time-dependence of $A_{\mathit{dc}}$.
The result is plotted in \fgref{fig:HEE_TimeDependence},
together with the $\gamma^2$-perturbation, the early time  and late time approximations
discussed in the last section.
The  figure shows that the $\gamma^2$-perturbation gives quite a good approximation
around $t_{\infty}\sim t_c$.

\vspace{0.1cm}

\paragraph{$\bm{\gamma^2 \to \frac{1}{2}}$}
When $\gamma^2$ is very close to $\frac{1}{2}$,
the time evolution of the entanglement entropy does not exhibit
a phase transition
for a large range of $\theta_{\infty}$.
The minimal value $\theta_c$ of the subsystem size $\theta_\infty$ necessary for the phase transition to happen  is determined by solving
\begin{align}
  A_{\mathit{dc}}(t_\infty=0,\theta_\infty=\theta_c,\gamma^2) = A_c(t_\infty=0,\gamma^2)\,.
\end{align}
By using the early-time expression \eqref{eq:earlyarea} for the left hand side,
we can solve this equation as%
\footnote{
We dropped subleading terms for $1-2\gamma^2$,
because 
in \eqref{eq:earlyarea}
we already used the $r_0\theta\gg 1$ approximation,
which in turn implies $1-2\gamma^2\ll 1$ here.
}
\begin{align}
  \theta_c 
\simeq \f{1}{2\s{2}r_0}\l(-\log{\l(1-2\gamma^2\r)}-2\log{2}\r)\,.
\end{align}
When $\theta\le\theta_c$,
the phase corresponding to the disconnected surface is realized 
from the initial time $t_{\infty}=0$.
Furthermore,
the initial entanglement entropy is proportional to the size of the subsystem ($\propto\theta$),
which can be also seen by using the early-time approximation.
This is one of the very peculiar point in the $\gamma^2 \to \frac{1}{2}$ limit.

\begin{figure}[htbp]
  \begin{center}
    \includegraphics[width=12cm]{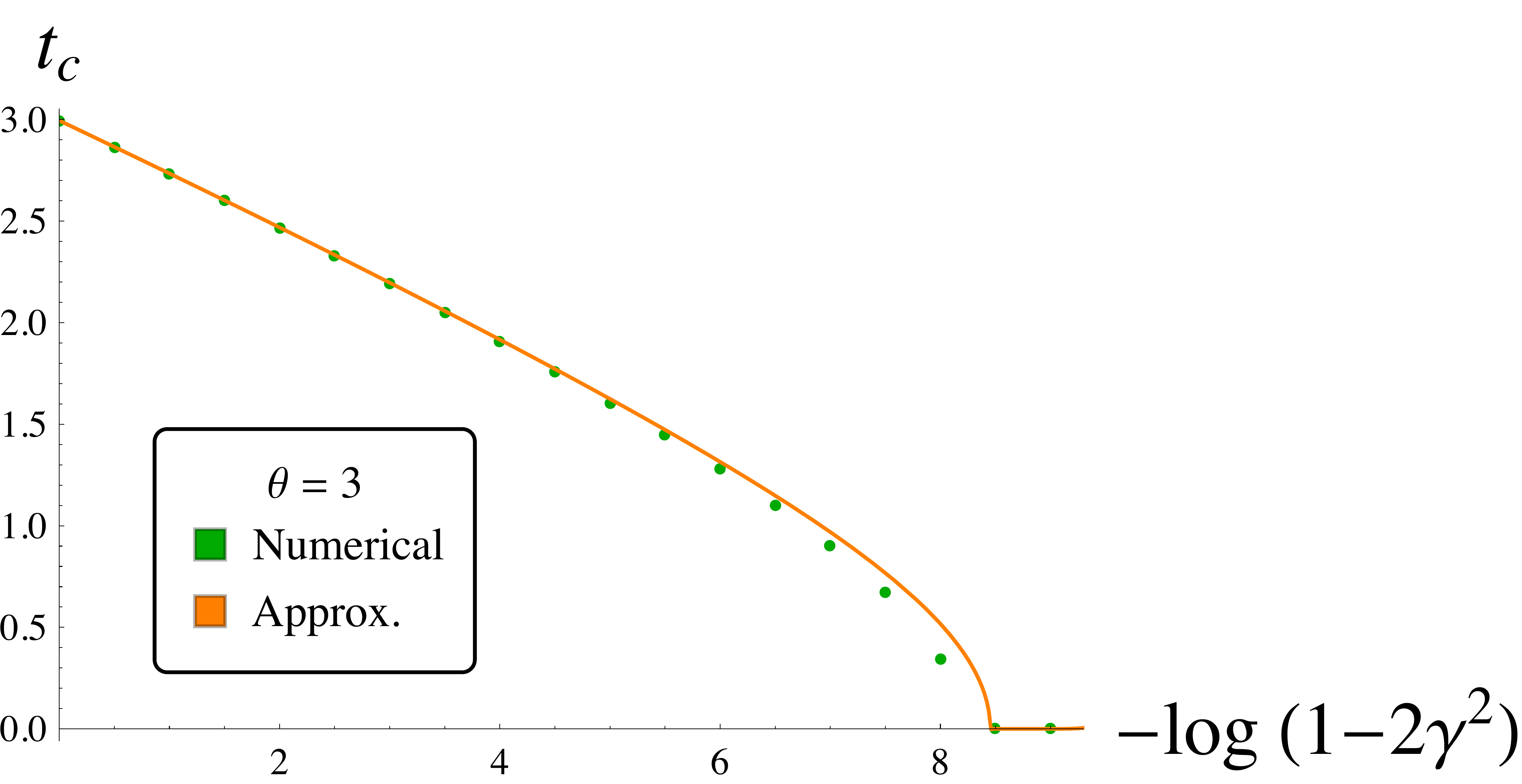}
  \end{center}
  \caption{The $\gamma^2$ dependence of the transition time $t_c$ of a subsystem
   $\theta=3$. The green dots are obtained by calculating the disconnected phase surface area $A_{\mathit{dc}}$ numerically. The approx line (orange line) is obtained by substituting the disconnected phase surface area $A_{\mathit{dc}}$ \eqref{eq:area:ren:expansion} calculated up to $\oo(\gamma^2)$. The transition time $t_c$ decreases with $-\log(1/2-\gamma^2)$ almost linearly, and the connected phase disappears with sufficiently large $\gamma^2$. }
  \label{fig:TransitionTime}
\end{figure}

In summary, although the time evolution of 
the holographic entanglement entropy 
in the Janus black hole is similar to that of the BTZ black hole, 
there are some significant differences.
First, 
the growth rate of the holographic entanglement entropy remains positive
even after the phase transition,
whereas the HEE for BTZ black hole is constant 
(i.e., the growth rate is zero) after $t_c$.
Second,
the introduction of the parameter $\gamma$
makes the connected surface less easy to realize.
It in turn brings a result that the transition time $t_c$
becomes earlier.
Accordingly,
the ``critical value'' $\theta_c$ for the subsystem size 
increases as $\gamma^2$ grows and approaches to $\f{1}{2}$.
Third, 
there is a nonzero initial entanglement entropy in general.
In particular,
when $\gamma^2$ is very close to $\frac{1}{2}$,
it is proportional to the size of the subsystem
even for relatively large $\theta_{\infty}$'s.
These results are hard to 
be understood in
the quasi-particle picture,
in contrast to the BTZ results.

\vspace{0.1cm}

It is also interesting to see the time evolution of the mutual information defined by 
\begin{align}
I(A;B)= S(A)+S(B)-S(A \cup B),
\end{align}
which measures the entanglement between two subsystems $A$ and $B$.
Here we take the subsystem $A$ to be an interval 
$-\theta_{\infty} < \theta < \theta_{\infty}$ in the right CFT, 
and $B$ to be the same interval in the left CFT. 
The original subsystem we have been considering is the union of them. 
Therefore the $I(A;B)$ eventually vanishes 
in the disconnected phase, $t_{\infty}\ge t_c$.
For BTZ black holes this 
critical time is given by half the size of the subsystem $t_c^{\mathrm{(BTZ)}}= \theta_{\infty}$ in the high temperature limit.    
In \cite{Shenker:2013pqa}, 
they considered the perturbation of BTZ black holes by a shock wave sent from one boundary, 
and found that the critical time becomes shorter by so called scrambling time. 
Here we see that 
our $\gamma$-deformation
also leads to earlier critical times.  
The main difference between our case and theirs is that 
the deviation of the critical time from the BTZ value $ t_{c}-t_{c}^{\mathrm{(BTZ)}}$ is proportional to the inverse temperature $\beta$ in their case, while it is not in our case.

\section{Conclusions}
\label{sec:conc}

\setlength{\parskip}{0.1cm}

In this paper, we considered a three-dimensional,
time-dependent two-sided black hole (Janus black hole) 
which can be regarded as a one parameter generalization of the BTZ black hole. 
This black hole contains a long wormhole region, which is causally disconnected from the conformal boundaries. 
The black hole is conjectured to be the dual of a particular CFT state (\ref{eq:cftstate}). 
The question here is how the information on the long wormhole region 
is encoded in the dual CFT state.

As a first step to answer this question, 
we calculated the time evolution of a holographic entanglement entropy $S_{A}$ in the black hole geometry, 
where the subsystem $A$ is the disjoint union of a region 
in the original CFT and a region in the thermofield double. 
In the calculation of the entropy, we considered the area of two 
(disconnected and connected)
extremal surfaces 
in the black hole geometry.

In  BTZ black hole geometry, the connected surface is initially chosen, 
then after the critical time $t_{c}$ which is proportional to the size of the subsystem, 
the disconnected surface becomes chosen. 
Although the behavior in the Janus black hole geometry shares many similarities to the BTZ case, 
there are two notable differences. 
First of all, we showed that the critical time is shorter than that in the BTZ case. 
Intuitively, this is because the Janus black hole has a longer wormhole region, 
therefore the length of the connected surface becomes longer than that of the BTZ black hole. 
We computed this critical time up to the second order of the deformation parameter $\gamma$. 
Secondly, we found that the disconnected surface is always chosen, 
when $\gamma^2$ gets sufficiently close to $1/2$
with the subsystem size fixed, namely, 
when the wormhole region is sufficiently long.

In \fgref{fig:SurfaceBarrier}, we numerically plotted the bulk region where the disconnected surface can arrive, 
and we found that outside the apparent horizon, there exists a barrier which any disconnected surface cannot go beyond.
As a result, after the phase transition, 
the black hole interior region that the entanglement entropy can probe
is rather limited.
This limitation is especially strong in the above case when $\gamma^2$ is close to $1/2$.

In \cite{Headrick:2014cta}, it was shown that if we take subsystem $A$ to be the total space of the  left CFT,
 the extremal surface which computes the holographic entanglement entropy has to
be located in the causal shadow.
This property  is necessary for the holographic entanglement entropy formula to respect the CFT causality. 
We can easily check this condition in the Janus black hole, because in the large $\theta_\infty$ limit 
the corresponding extremal surface localizes at the origin $(y,t)=(0,0)$ (or $(\mu, \tau)=(0,0)$ in the coordinate \eqref{eq:metric:tau}).

There are several outlooks for this work. 
It would be interesting to calculate the entanglement entropy on the dual CFT side. 
One candidate CFT is a free fermion system \cite{Morrison:2012iz}, 
for which the explicit form of the twist operator is known \cite{Azeyanagi:2007bj}. 
\fgref{fig:SurfaceBarrier} seems to show that 
it is not possible for the disconnected surface to penetrate the apparent horizon
of the Janus black hole,
and
it would be interesting to prove this 
directly like \cite{Engelhardt:2013tra}.

\setlength{\parskip}{0cm}

\section*{Acknowledgments}
\addcontentsline{toc}{section}{\protect\numberline{}Acknowledgments}
The authors thank 
Matthew Headrick,
Tatsuma Nishioka
and 
Tadashi Takayanagi 
for fruitful discussions. 
The work of Y.\,N. was supported
by Japan Society for the Promotion of Science (JSPS) 
Research Fellowship for Young Scientists,
in part
by JSPS Grant-in-Aid for JSPS Fellows,
and also in part
by World Premier International Research Center Initiative
(WPI) from
Ministry of Education, Culture, Sports, Science and Technology 
(MEXT) of Japan.
The work of N.\,O. was supported by the Special Postdoctoral
Researcher (SPDR) Program of RIKEN
and in part
by the interdisciplinary Theoretical Science (iTHES) Project of RIKEN.
The work of T.\,U. was supported in part by the National Science Foundation under Grant No.\,NSF\, PHY-25915. 
The work of T.\,U. was supported
by JSPS Postdoctoral Research Fellowship for Young Scientists
and in part
by JSPS Grant-in-Aid for JSPS Fellows,
in the earlier stage of this work.

\appendix
\section{The $\gamma$-expansion of Holographic Entanglement Entropy}
\label{sec:gamma_expand}
In this section, we compute the entanglement entropy and the phase transition time in the leading order of $\gamma $-expansion.

\subsubsection*{Expansion of $y_*$}
First, let us expand \eqref{eq:eom:ystar}.
The integrand in the left hand side is expanded as
\begin{align}
  \frac{\ti{g}(y)^2}{\sqrt{\ti{g}(y_*)^2-\ti{g}(y)^2}}
  &=
  \f{\sech^2{y}}{\s{\sech^2{y_*}-\sech^2{y}}}
  + \f{\sech^2{y}(2-\sech^2{y}+\sech^2{y_*})}{4\s{\sech^2{y_*}-\sech^2{y}}}\gamma^2
  +\mathcal{O}(\gamma^4)\,,
\end{align}
then 
\begin{align}
  &\int_{y_*}^{y_\infty}dy\,\frac{\ti{g}(y)^2}{\sqrt{\ti{g}(y_*)^2-\ti{g}(y)^2}}
\nonumber\\
&\quad= 
\l[\tanh^{-1}\l(\f{\cosh{y_*}\sinh{y}}{\s{\cosh^2{y_*}-\cosh^2{y}}}\r)\r]_{y_*}^{y_\infty}
\nonumber\\
&\qquad+ \l[\f{3\cosh^2{y_*}+1}{8\cosh^2{y_*}}\tanh^{-1}\l(\f{\cosh{y_*}\sinh{y}}{\s{\cosh^2{y_*}-\cosh^2{y}}}\r)
+ \f{1}{8\cosh{y_*}}\f{\s{\cosh^2{y}-\cosh^2{y_*}}}{\sinh{y}}
\r]_{y_*}^{y_\infty}\gamma^2
+ \mathcal{O}(\gamma^4)
\nonumber\\
&\quad= 
\tanh^{-1}(\sech{y_*}) + \l(\f{3\cosh^2{y_*}+1}{8\cosh^2{y_*}}\tanh^{-1}(\sech{y_*})+ \f{1}{8\cosh{y_*}}\r)\gamma^2 + \mathcal{O}(\gamma^4)\,.
\end{align}
By substituting this into \eqref{eq:eom:ystar}, we obtain
\begin{align}
  \f{\sinh{r_0\theta}}{\cosh{r_0t}}
&= \sinh\l[
\tanh^{-1}(\sech{y_*}) + \l(\f{3\cosh^2{y_*}+1}{8\cosh^2{y_*}}\tanh^{-1}(\sech{y_*})+ \f{1}{8\cosh{y_*}}\r)\gamma^2 + \mathcal{O}(\gamma^4)
\r]
\nonumber\\
&= 
\f{1}{\sinh{y_*}}
+ \l(\f{3\cosh^2{y_*}+1}{8\cosh{y_*}\sinh{y_*}}\tanh^{-1}(\sech{y_*})+ \f{1}{8\sinh{y_*}}\r)\gamma^2
+ \mathcal{O}(\gamma^4)\,,
\end{align}
leading to
\begin{align}
  \sinh{y_*} = F \l[ 1 + 
\l(
\f{3F^2+4}{8\s{1+F^2}}\coth^{-1}\l(\s{1+F^2}\r) + \f{1}{8}
\r)\gamma^2 \r] + \mathcal{O}(\gamma^4)\,,
\label{eq:ystar:expansion}
\end{align}
where 
\begin{align}
  F(t,\theta) = \f{\cosh{r_0t}}{\sinh{r_0\theta}}\,.
\end{align}

\subsubsection*{Disconnected surface area}
On the other hand, from \eqref{eq:Adc:y} and \eqref{eq:area:ren},
$\gamma^2$-expansion gives
\begin{align}
  A_{\mathit{dc}}^{\mathit{(ren)}}/L
&=
  4\log\f{2\cosh{r_0t}}{r_0\sinh{y_*}}
+ \l(2+ \sech{y_*}\tanh^{-1}(\sech{y_*})\r)\gamma^2
+ \mathcal{O}(\gamma^4)\,.
\end{align}
By using \eqref{eq:ystar:expansion} above, this results in
\begin{align}
  A_{\mathit{dc}}^{\mathit{(ren)}}/L
&=
  4\log\l(\f{2}{r_0}\sinh{r_0\theta}\r)
- \l(\f{3F^2+2}{2\s{1+F^2}}\coth^{-1}\l(\s{1+F^2}\r)-\f{3}{2}\r)\gamma^2 + \mathcal{O}(\gamma^4)\,.
\label{eq:area:ren:expansion:appendix}
\end{align}

\subsubsection*{Phase Transition}
The phase transition time $t_c$ for a fixed value of $\theta$ can be computed by an equation
\begin{align}
  A_{\mathit{dc}}^{\mathit{(ren)}} = A_c^{\mathit{(ren)}}\,,
\end{align}
with the aid of 
\eqref{eq:Ac:ren} and \eqref{eq:area:ren:expansion:appendix}.
This equation is solved as $t=t_c$, where
\begin{align}
  t_c &= t_c^{(0)} + t_c^{(1)}\gamma^2 + \mathcal{O}(\gamma^4)\,,\\
  r_0 t_c^{(0)} &= \cosh^{-1}(\sinh{r_0\theta})\,, \\
  r_0 t_c^{(1)} &= - \l(\f{1}{2} + \f{5}{2\s{2}}\coth^{-1}(\s{2})\r)\f{\sinh{r_0\theta}}{\s{\sinh^2{r_0\theta}-1}}\nonumber\\
  &\simeq -2.058\times\f{\sinh{r_0\theta}}{\s{\sinh^2{r_0\theta}-1}}\,.
\end{align}
Then, in particular, in the large $\theta$ limit ($\theta \gg r_0^{-1}$),
we obtain 
\begin{align}
  t_c \simeq \theta - 2.058\gamma^2 + \mathcal{O}(\gamma^4)\,.
\end{align}

\bibliographystyle{JHEP}
\bibliography{reference}

\end{document}